\documentclass[doublecol,figures]{epl2} 
\usepackage{amssymb,amsmath}
\usepackage{url}

\newcommand{\bea}{\begin{eqnarray}}
\newcommand{\eea}{\end{eqnarray}}
\newcommand{\be}{\begin{equation}}
\newcommand{\ee}{\end{equation}}
\newcommand{\kk}{\mathbf{k}}

\newcommand{\OO}{\mathbf{0}}
\newcommand{\rr}{\mathbf{r}}
\newcommand{\pp}{\mathbf{p}}

\newcommand{\zero}{\mathbf{0}}

\title{Limit of spin squeezing in trapped Bose-Einstein condensates}
\shorttitle{Limit of spin squeezing in trapped Bose-Einstein condensates}
\author{Alice Sinatra$^1$, Yvan Castin$^1$ and Emilia Witkowska$^2$}  
\shortauthor{Alice Sinatra, Yvan Castin and Emilia Witkowska}
\institute{$^1$Laboratoire Kastler Brossel, \'Ecole Normale Sup\'erieure and CNRS, UPMC, 24 rue Lhomond, 75231 Paris, 
France \\
$^2$Institute of Physics, Polish Academy of Sciences,
Aleja Lotnik\'ow 32/46, 02-668 Warszawa, Poland}


\abstract{
The evolution of an interacting two-component Bose-Einstein condensate from an initial phase state leads
to a spin squeezed state that may be used in atomic clocks to increase the signal-to-noise ratio,
opening the way to quantum metrology.
The efficiency of spin squeezing is limited by the finite temperature of the gas, as was 
shown theoretically
in a spatially homogeneous system. Here we determine the limit of spin squeezing
in the realistic trapped case,  with classical field simulations,
and with a completely analytical treatment that includes the quantum case. }

\begin{document}
\maketitle
\date{\today}

\section{Introduction}
In good atomic clocks, the signal-to-noise ratio is determined by the quantum noise, that is the partition
noise of uncorrelated atoms among two internal levels $a$ and $b$, 
rather than by technical noise \cite{Santarelli}. 
The resulting statistical uncertainty on the transition frequency $\omega_{ab}$ is then $\Delta \omega_{ab}
=1/(\tau N^{1/2})$ where $N$ is the atom number and $\tau$ the Ramsey interrogation time.

One can beat this so-called ``standard quantum limit" by introducing appropriate quantum correlations
among the atoms, that is using squeezed states of the collective spin \cite{Ueda}
having reduced fluctuations of standard deviation $\Delta S_\perp$ along some direction 
transverse to the mean spin 
$\langle \mathbf{S}\rangle$. The resulting statistical uncertainty on the transition frequency
is reduced by the squeezing parameter $\xi < 1$ to the value \cite{Wineland}
\be
\label{eq:wine}
\Delta \omega_{ab}^{\rm sq} = \frac{\xi}{N^{1/2} \tau} \ \ \mbox{with}\ \ \xi^2 = \frac{N \Delta S_\perp^2}
{|\langle \mathbf{S}\rangle|^2}.
\ee

A fundamental issue is then to know how $\xi$ scales in the large-$N$ limit. 
A practical realisation of spin squeezing requires some non-linearity in the spin dynamics 
\cite{Polzik,Vuletic}. 
In two-component Bose-Einstein condensates, it is provided by the atomic $s$-wave interactions \cite{NatureSorensen},
and first experimental implementations have been performed with a dynamical control
of the interactions \cite{manips}. 
The original model Hamiltonian proposed in \cite{Ueda} is then realized if one restricts to the two condensate
spatial modes (in {hyperfine states} $a$ and $b$), resulting in a squeezing parameter minimised over time
such that $\xi_{\rm min}^2 \approx N^{-2/3}$. For one million atoms, this predicts a signal-to-noise
increase by a factor $100$. There may be however serious shortcomings of the two-mode model.
The combined effect of one-body and three-body particle losses was shown in \cite{LiYun} 
to impose a non-zero lower bound 
on $\xi_{\rm min}$. More fundamentally, the finite temperature gases in the experiments are intrinsically
multimode; in the spatially homogeneous case, this was predicted in \cite{prlsqueezing,italien} to lead to a
non-zero value of $\xi^2_{\rm min}$ in the thermodynamic limit, in sharp contrast with the
$N^{-2/3}$ scaling, as a consequence of a random dephasing of the condensate introduced by 
the non-condensed quasi-particles \cite{frontiers}.

Here, we perform the last step towards a full theoretical understanding of the spin squeezing
dynamics, by including the effect of the trap present in the experiments. Although
it significantly complicates the theoretical analysis, we will see that an analytical treatment
can be pushed to the end to obtain $\xi_{\rm min}$ in the
large system-size limit.

\section{Classical fields}
\label{sec:themodel}

\begin{figure}[t]
\includegraphics[width=\columnwidth,clip=]{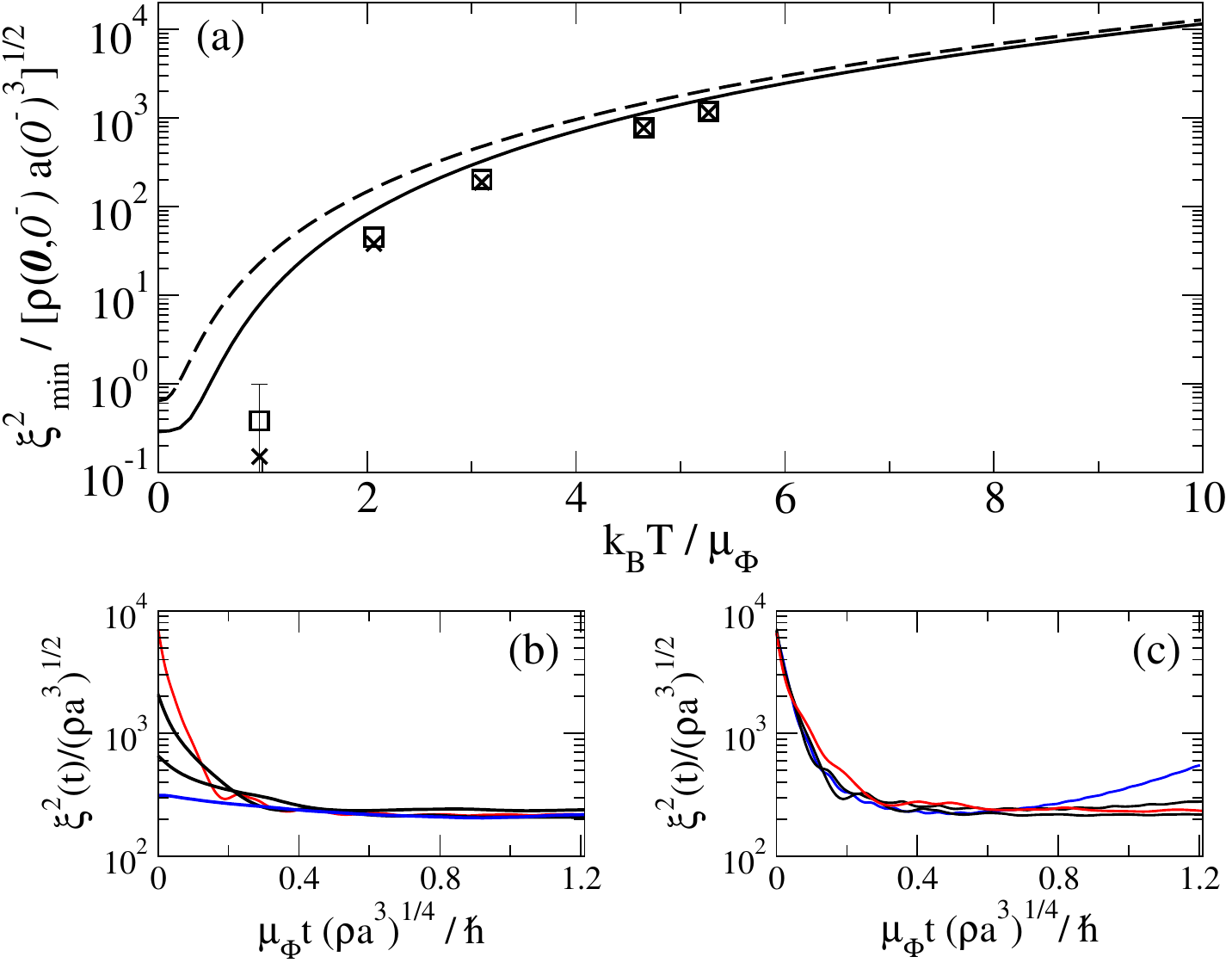}
\caption{Top (a): Minimal spin squeezing parameter $\xi_{\rm min}^2$ in an isotropic harmonic trap,
$U(\rr)=m\omega^2 r^2/2$, as a function of temperature. 
Quantum fields: analytical result (\ref{eq:ana_quant}) (solid line);
{local density approximation (dashed line)}.
Classical field model: simulations (squares) vs analytics (\ref{eq:ana_class}) (crosses).
{Bottom: For $k_BT/\mu_{\Phi}=2.89$, rescaled time-dependent $\xi^2(t)$  (b) for  
$[\rho(\zero,0^-) a^3(0^-)]^{1/2}$ between $3\times10^{-3}$ (blue) and $5\times10^{-4}$ (red) with
$\mu_\Phi/\hbar \omega \simeq 5.1$ and
(c) for $N$ between $3.7\times10^3$ (blue) and $5.6\times10^5$ (red) with
$[\rho(\zero,0^-) a^3(0^-)]^{1/2}=1.4\times 10^{-4}$.}
\label{fig:xi2class}}
\end{figure}

Classical field models require an energy cut-off of the order of $k_B T$, where $T$ is the initial
temperature of the gas. Here it is provided by a cubic lattice model, with a unit cell volume $dV$ 
adjusted to reproduce the quantum non-condensed particle number for a zero chemical potential
spatially homogeneous ideal Bose gas. The corresponding Hamiltonian is
\be
H=\sum_{\sigma=a,b} \sum_\rr  dV \left[\psi_\sigma^* h_0 \psi_\sigma 
+\frac{g(t)}{2} \psi_\sigma^* \psi_\sigma^* \psi_\sigma \psi_\sigma\right]
\ee
with a one-body part $h_0=\pp^2/(2m) + U(\rr)$ involving the kinetic energy part [eigenmodes 
are plane waves of wavevector $\kk$ in the first Brillouin zone] and the internal-state
independent trapping part $U(\rr)$. 
We assume that the coupling constant $g(t)$ is the same in the two internal states $a$ and $b$, as for $F=1$ spinor condensates, and that there is no interaction between
$a$ and $b$, which may be realized by spatial separation of the two components.
Initially, at time $t=0^-$, the gas is in thermal equilibrium in the internal state
$a$, with a fixed total particle number $N$ and a coupling constant $g(0^-)$. 
To mimic quantum noise in the
initially empty internal state $b$, we sample the Gaussian Wigner distribution of the vacuum state in $b$,
which amounts to introducing on average $1/2$ fictitious particles in 
each of the $\mathcal{N}$ spatial
modes\footnote{The Wigner method in spin squeezing context
was also used in \cite{Isella} and recently extended to the case with particle losses
\cite{Opanchuk}.}.
At $t=0$, the gas experiences a $\pi/2$ pulse, inducing 
the field mapping\footnote{{We consider here a microwave-transition scheme.  
As other schemes, e.g.\ cavity-based \cite{Lukin}, it is interesting also for optical clocks.}}
\be
\label{eq:pulse}
\psi_{a,b}(\rr,0^+)= \frac{1}{\sqrt{2}}[\psi_a(\rr,0^-) \mp \psi_b(\rr,0^-)]
\ee
This distributes on the average half of the particles in each internal state, with Poissonian
fluctuations, and it puts the system out of equilibrium. To minimize the excitation of a breathing
mode of the condensate \cite{spsq_dyn}, we subsequently increase the coupling constant by a factor two,
\be
\label{eq:trick}
g(0^+) \equiv g = 2 g(0^-),
\ee
which may be realized using a Feshbach resonance
\footnote{Strictly speaking, one should take $g(0^-) N=g(0^+) (N+\mathcal{N})/2$. Here, however, 
one has $N\gg \mathcal{N}/2$ as required in \cite{Cartago}.}
\footnote{ We have however seen numerically that for $g(0^+)=g(0^-)$ such excitation does not
have a large impact on the squeezing.}. 
The before-pulse coupling constant $g(0^-)$
is related to the before-pulse $s$-wave scattering length by $g(0^-)=4\pi\hbar^2a(0^-)/m$.
After the pulse, each field $\psi_{a,b}$ is evolved according to the non-linear Schr\"odinger equation
\be
\label{eq:motion}
i\hbar \partial_t \psi_\sigma = [h_0 + g |\psi_\sigma(\rr,t)|^2] \psi_\sigma
\ee
The variances of the various collective spin components 
$S_x+iS_y=\sum_\rr dV \psi_a^*(\rr) \psi_b(\rr)$ and $S_z=(N_a-N_b)/2$, where $N_\sigma$ is the particle number 
in internal state $\sigma$, give access to the spin squeezing
parameter $\xi^2(t)$ as a function of time. According to Eq.~(\ref{eq:wine}), one needs in
particular \cite{prlsqueezing}:
\be
\Delta S_\perp^2 = \frac{1}{2} [\langle S_y^2\rangle + \langle S_z^2\rangle -|\langle(S_y+iS_z)^2\rangle|],
\ee
where $\langle \ldots\rangle$ represents the average over all stochastic realisations.
In practice, spin squeezing is supposed to be applied to large atomic ensembles,
as in clocks, so we concentrate our study on the thermodynamic limit,
where the single-particle level spacing tends to zero for fixed temperature and chemical potential. 
We then find as in \cite{prlsqueezing,italien} 
that $\xi^2(t)$ is a very flat function around its minimum, 
and that its minimum tends
to a finite value $\xi_{\rm min}^2$ in the thermodynamic limit, {see
fig.~\ref{fig:xi2class}c}. This is a striking consequence
of the multimode nature of the problem.

In the spatially homogeneous case, in the weakly interacting limit,
$\xi_{\rm min}^2$ obeys a simple scaling formula \cite{prlsqueezing,italien}. Its intuitive generalisation to the trapped
case is
\be
\frac{\xi_{\rm min}^2}{[\rho(\OO,0^-)a^3(0^-)]^{1/2}} = f\left(\frac{k_B T}{\mu_{\Phi}}\right) \label{eq:sca}
\ee
where $\mu_{\Phi}$ is the Gross-Pitaevskii chemical potential of the gas, and $\rho(\OO,0^-)$ is the mean 
atomic density before the pulse in the trap center.
{We have checked numerically
that this scaling holds for the trapped system, see fig.~\ref{fig:xi2class}b,
which also shows that the different curves plateau 
at a common rescaled time, as in \cite{prlsqueezing,italien}.}
The rescaled minimal spin squeezing $\xi_{\rm min}^2$ as a function of the rescaled temperature,
that is the function $f$, is plotted with squares in fig.~\ref{fig:xi2class}a.
The temperature is limited to the range $k_B T > \mu_{\Phi}$ where our classical field model makes sense. 

\section{Analytical approach for classical fields}
We develop analytics within the framework of the number-conserving Bogoliubov approach \cite{
Gardiner,CastinDum},
along the lines of our studies of the homogeneous case \cite{italien}.
The Bogoliubov approach is used first to describe the initial distribution of the fields (before the pulse);
this is relatively standard, and we shall be brief on this aspect here. 
Second, it is used to describe the dynamical evolution of the fields after the pulse, which is more
involved as we now explain.

The theory introduces as a small parameter the non-condensed fraction of the gas,
$\epsilon_{\rm Bog}$, and it is pushed to first order included in $\epsilon_{\rm Bog}$.
It would remain quite involved without the use of a second small
parameter, 
$\epsilon_{\rm size} = 1/N^{1/2}$ where $N$ is the total particle number, 
controlling the approach of the thermodynamic limit 
\footnote{The two small parameters
are not fully independent. Since the system size is $\gg$ the thermal wavelength, one finds
$N\epsilon_{\rm Bog} \gg 1$.}.
In the Bogoliubov approach, one singles out the contribution of the macroscopically 
populated modes, corresponding to the condensate wavefunctions $\phi_\sigma$ in each internal state:
\be
\psi_\sigma(\rr,t) = a_\sigma(t) \phi_\sigma(\rr,t) + \psi_{\perp\sigma}(\rr,t)
\label{eq:splitting}
\ee
The non-condensed fields $\psi_{\perp \sigma}$ are orthogonal to the condensate wavefunctions; their squared modulus
gives the non-condensed density so that they are smaller than the condensate fields
$a_\sigma  \phi_\sigma$ by a factor $\epsilon_{\rm Bog}^{1/2}$. This allows the Bogoliubov  expansion
of the Hamiltonian in powers of $\psi_\perp$.

For each component $\sigma$ the condensate wavefunction $\phi_\sigma$ is an averaged quantity: 
it is the most populated eigenstate of the one-body density matrix
$\langle{\psi^*_{\sigma}(\rr',t) \psi_\sigma(\rr,t)}\rangle$.
Due to symmetry in our problem, one has 
$\phi_a=\phi_b$. The expansion of $\phi_\sigma$ in powers of $\epsilon_{\rm Bog}^{1/2}$ reads, with
notations slightly different from \cite{CastinDum}:
\be
\label{eq:dev}
\phi_\sigma(\rr,t) = \Phi(\rr) + \phi^{(2)}(\rr,t) + O\left[(\epsilon_{\rm Bog}^{1/2})^3\right]
\ee
The zeroth order contribution $\Phi$ corresponds to Gross-Pitaevskii theory; it is time-independent here,
due to the trick (\ref{eq:trick}), so it can be taken as the real and positive solution of 
the stationary Gross-Pitaevskii equation
\be
\label{eq:gpe}
0 = [h_0 + g \langle N_\sigma\rangle  \Phi^2-\mu_{\Phi}]\Phi
\ee
where $\langle N_\sigma\rangle$ is the mean particle number in each internal state after
the pulse. The equation
for the correction $\phi^{(2)}$, due to condensate depletion and interaction with 
the non-condensed particles, is not needed; we shall use
that $\phi^{(2)}=O(\epsilon_{\rm Bog}\Phi)$ and can be 
taken orthogonal to $\Phi$ \cite{CastinDum}: {this sets the phase of $\phi_\sigma$
and restricts the phase evolution to $a_\sigma$}.

Of crucial importance is the modulus-phase representation of the condensate amplitudes
\be
a_\sigma(t) = e^{i\theta_\sigma(t)} [n_\sigma(t)]^{1/2}
\ee
where the condensate phases $\theta_\sigma$ and particle numbers $n_\sigma$ are fluctuating and time-dependent
quantities. First, inclusion of this phase factor in the non-condensed fields leads to the
main dynamical variables of the number-conserving theories:
\be
\Lambda_\sigma(\rr,t) \equiv e^{-i\theta_\sigma(t)} \psi_{\perp\sigma}(\rr,t)
\ee
Second, the phases $\theta_\sigma(t)$ increase linearly in time,
contrarily to the condensate numbers $n_\sigma(t)$ or the fields $\Lambda_\sigma$ that are bounded
in time. The simulations show that the close-to-best squeezing time remains finite in the thermodynamic limit,
as it was the case for the spatially homogeneous system \cite{prlsqueezing,italien};
the phase difference $(\theta_a-\theta_b)(t)$
thus remains $O(1/N^{1/2})$ and $\ll 1$ over the relevant time window.
Linearizing the quantity $e^{i(\theta_b-\theta_a)}$ that appears in the expression of the collective
spin component $S_y$, as in \cite{prlsqueezing,italien}, 
one obtains a contribution growing linearly in time that eventually
dominates the collective spin component $S_z$ (which is a constant of motion).
The resulting squeezing parameter $\xi^2(t)$ reaches its minimal value at large times 
\footnote{ As in the homogeneous case, large times correspond (in the Thomas-Fermi regime) to
$\xi_{\rm min} \mu_{\Phi} t/\hbar \gg 1$.} \footnote{As in \cite{italien} we
use the fact that, at large times, $S_y\simeq (\theta_b-\theta_a) \mathcal{F}$
where $\mathcal{F}=\sqrt{n_an_b}+\mbox{Re}\, \sum_\rr dV \Lambda_a^* \Lambda_b$ can be replaced by its mean
value $\langle \mathcal{F}\rangle$ in the thermodynamic limit since $\theta_a-\theta_b$ is of zero mean.}:
\be
\xi^2(t) \underset{\mathrm{large}\ t}{\simeq } 
1-\frac{\langle(\theta_a-\theta_b)(N_a-N_b)\rangle^2}
{\langle(\theta_a-\theta_b)^2\rangle \langle (N_a-N_b)^2\rangle}
\underset{\mathrm{large}\ t}{\simeq } \xi_{\rm min}^2 
\label{eq:asympt}
\ee
Note that $N_a-N_b$ scales as $N^{1/2}$, since it is of zero mean and of variance $\simeq N$.

We thus need to evaluate the time evolution of the condensate phases. From the equation of motion 
(\ref{eq:motion}), one gets the expression of $\frac{d}{dt}\theta_\sigma$ that we transform
with the splitting (\ref{eq:splitting}), keeping terms up to order one included in $\epsilon_{\rm Bog}$ 
and up to order one included in $\epsilon_{\rm size}$, and using where necessary
$n_\sigma=N_\sigma-\sum_\rr dV |\Lambda_\sigma|^2$. Going beyond first order in $\epsilon_{\rm Bog}$
is indeed beyond Bogoliubov theory. According to (\ref{eq:asympt}), the relevant quantity
is $N^{1/2}(\theta_a-\theta_b)$, and terms in the phase difference that are $o(\epsilon_{\rm size})$ 
are negligible in the thermodynamic limit.  Also, the crossed terms 
scaling as $\epsilon_{\rm Bog}\epsilon_{\rm size}$ are negligible.
{Combining the two transverse corrections to the pure-condensate Gross-Pitaevskii field
\be
\Xi_\sigma(\rr,t) \equiv \Lambda_\sigma(\rr,t) + \langle N_\sigma\rangle^{1/2} \phi^{(2)}(\rr,t)
\label{eq:Xi}
\ee
we finally obtain after the pulse
\begin{multline}
-\hbar \frac{d \theta_\sigma}{dt}  \simeq   \mu_{\Phi} +    \left( \langle N_\sigma\rangle^{1/2} + q_\sigma \right) g 
\!\int\! \Phi^3 \left( \Xi_\sigma+\mbox{c.c.} \right)  \\
+ (n_\sigma-\langle N_\sigma\rangle) g \!\int\! \Phi^4 + 
2 g \!\int\! \Phi^2 \left(|\Lambda_\sigma|^2 +\frac{\Lambda_\sigma^2+\mbox{c.c.}}{4}\right)
\label{eq:derivtheta}
\end{multline}}
where $\int$ is a short-hand notation for $\sum_\rr dV$ and we have introduced
the quantities of order unity corresponding to the fluctuations in the particle numbers,
$q_\sigma \equiv (N_\sigma-\langle N_\sigma\rangle)/\langle N_\sigma\rangle^{1/2}$.
{The mean field contribution $\mu_{\Phi}$ and the second line in (\ref{eq:derivtheta}) were already present in the spatially homogeneous case \cite{phase1}. 
The $\Xi_\sigma$ term in (\ref{eq:derivtheta}) that is linear in $\Lambda_\sigma$ is new and 
potentially larger by a factor $\epsilon_{\rm Bog}^{-1/2}$
than the second line.} It forces us to evaluate the fields $\Lambda_\sigma$ one order in $\epsilon_{\rm Bog}^{1/2}$
beyond the usual Bogoliubov approximation, which makes the trapped case treatment more subtle than the
homogeneous case, as we now see.

In the usual number-conserving Bogoliubov approaches, the fields $\Lambda_\sigma$ are evaluated to 
the leading $\epsilon_{\rm Bog}^{1/2}$ order only. They are expanded on the Bogoliubov mode functions
$u_k(\rr)$ and $v_k(\rr)$, that are orthogonal to $\Phi$ and normalized as
$\langle u_k|u_k\rangle - \langle v_k | v_k\rangle = 1$. They are here common to both internal states
and most importantly, 
thanks to the trick (\ref{eq:trick}), they are at $t<0$ and $t>0$ the same
time-independent eigenstates of the Bogoliubov operators of eigenenergies $\epsilon_k >0$.
To first order included in $\epsilon_{\rm Bog}^{1/2}$,
the modal expansion then reads (after the pulse)
\be
\label{eq:usual}
\Lambda_\sigma^{(1)}(\rr,t) = \sum_k c_{k\sigma}(0^+) e^{-i\epsilon_k t/\hbar} u_k(\rr) 
+  c_{k\sigma}^{*}(0^+) e^{i\epsilon_k t/\hbar} v_k^*(\rr)
\ee
From (\ref{eq:pulse}) and along the lines of \cite{italien}, one finds
\be
\label{eq:init}
c_{k a,b}(0^+) = \frac{1}{\sqrt{2}} [c_{ka}(0^-)\mp B_k],
\ee
omitting a $O(1/N^{1/2})$  contribution involving the condensate phase change due the pulse, 
negligible
in the large $N$ limit \cite{italien}. The modal coefficients $c_{ka}(0^-)$ in the initial thermal equilibrium state
have independent Gaussian distributions, with $\langle |c_{ka}(0^-)|^2\rangle = k_B T/\epsilon_k$ 
according to the equipartition
formula. The amplitudes $B_k$ result from the projection of the initial Wigner noise in internal state $b$ over
the Bogoliubov modes, and are thus Gaussian variables of zero mean statistically independent from the 
$c_{ka}(0^-)$'s. We only need here that 
$\langle |B_k|^2\rangle = \frac{1}{2} + \langle v_k| v_k\rangle$.

Due to the potentially much larger second term in Eq.~(\ref{eq:derivtheta}), the leading order 
(\ref{eq:usual}) is not accurate enough, and one must go to the next order in $\epsilon_{\rm Bog}^{1/2}$.
Furthermore the fluctuations of the particle numbers $N_\sigma$ will give a contribution not included
in (\ref{eq:usual}), since the usual number-conserving theory assumes a fixed total particle
number. One writes $\Lambda_\sigma= e^{-i\theta_\sigma} Q_\phi \psi_\sigma$, where 
$Q_\phi$ projects orthogonally to $\phi_\sigma$, and one calculates the temporal derivative.
After insertion of the splitting (\ref{eq:splitting}) and of the expansion (\ref{eq:dev}), 
one gets terms scaling as $\mu_{\Phi} \Lambda_\sigma 
\epsilon_{\rm Bog}^\alpha \epsilon_{\rm size}^\beta$; such terms contribute to the second term
of Eq.~(\ref{eq:derivtheta}) as $\mu_{\Phi} \epsilon_{\rm Bog}^{\alpha+1/2}\epsilon_{\rm size}^\beta$, so they have to be 
neglected if $\alpha+1/2 >1$ or $\beta>1$ or if $\alpha+1/2=\beta=1$. 
{It turns out  that the combined field (\ref{eq:Xi}) naturally appears\footnote{{For fixed $N_\sigma$, $\Lambda_\sigma$ differs from the zero-mean field $\Lambda_{\rm ex}$
of \cite{CastinDum} by $O(\epsilon_{\rm Bog})\Lambda$. Then the expectation value of Eq.(\ref{eq:derivlambda}) correctly gives the classical-field version of Eqs.(95,96) for $\phi^{(2)}$ in
\cite{CastinDum}.}}:
\begin{multline}
\label{eq:derivlambda}
i\hbar \partial_t \Xi_\sigma \simeq  Q (h_0-\mu_{\Phi} +2 \langle N_\sigma\rangle g \Phi^2) \Xi_\sigma 
+ Q\langle N_\sigma\rangle g \Phi^2 \Xi_\sigma^*  \\
+ q_\sigma \langle N_\sigma\rangle^{1/2} [Q g\Phi^2(2\Lambda_\sigma+\Lambda_\sigma^*) -
g\Lambda_\sigma \!\!\int\!\! \Phi^4 ]  \\
+ (n_\sigma-\langle N_\sigma\rangle) \langle N_\sigma\rangle^{1/2} Q g \Phi^3  \\
+ \langle N_\sigma\rangle^{1/2} g Q \{\Phi (2 |\Lambda_\sigma|^2+\Lambda_\sigma^2)-\Lambda_\sigma
\!\int\! [\Phi^3 (\Lambda_\sigma+\Lambda_\sigma^*)]\}
\end{multline}}
where $Q$ projects orthogonally to $\Phi$. The various terms in the right-hand side
have a simple physical interpretation. The first two terms constitute the usual Bogoliubov equations of motion, written
for a number $\langle N_\sigma\rangle$ of particles, and of which
$(u_k,v_k)$ are the positive-energy eigenmodes. Due to the fluctuations of the particle numbers, 
these usual Bogoliubov equations differ from the ones written for a given realisation with
$N_\sigma$ particles, hence the {third} term.
The {fourth} term corresponds to a similar effect at the level of the 
Gross-Pitaevskii equation, further including the condensate depletion: It takes into account the fact that the Gross-Pitaevskii solution for
$n_\sigma$ particles in the condensate of internal state $\sigma$ 
slightly differs from the solution $\Phi$ for $\langle N_\sigma\rangle$ particles in that condensate, providing
a source term for $\Lambda_\sigma$.
This fourth term already appeared in Eq.~(3.14) of \cite{Sorensen} (except for the missing projector $Q$).
The {fifth} term describes the effect of the cubic interaction among the Bogoliubov quasi-particles;
whereas this interaction usually leads to a true transfer of quasi-particles among the Bogoliubov
modes (according to the so-called Beliaev-Landau processes), its relevant effect here is rather a reactive
effect that shifts the value of the field $\Lambda_\sigma$.

We now apparently have to solve Eq.~(\ref{eq:derivlambda}). This formidable task is greatly simplified
by the fact that we need here, in the long time limit, only the linearly diverging part of the phases
$\theta_\sigma(t)$. What should be evaluated is thus only the time averaged part 
$\bar{\theta}_\sigma$; the oscillating parts $\theta_\sigma(t)-\bar{\theta}_\sigma$ will be neglected
in the so-called {\sl secular} approximation for the condensate phase \cite{phase1} carefully
justified for spin squeezing in the homogeneous case in \cite{italien}.
In the second line of (\ref{eq:derivtheta}), the leading value (\ref{eq:usual}) for
$\Lambda_\sigma$ suffices. In the second term of (\ref{eq:derivtheta}), it would give
the insufficient result $\bar{\Lambda}_\sigma=0$, so we have to determine the leading order
non-zero approximation for $\bar{\Lambda}_\sigma$ from (\ref{eq:derivlambda}).
In the subleading terms of (\ref{eq:derivlambda}), that is the third, fourth and fifth terms,
we can replace $\Lambda_\sigma$ by the leading value (\ref{eq:usual}); after temporal average,
the third term thus disappears. Adding the temporally averaged (\ref{eq:derivlambda}) to its
complex conjugate gives the inhomogeneous equation
\be
0=M (\bar{\Xi}_\sigma+\bar{\Xi}_\sigma^*)+S
\ee
with  the {\sl hermitian} operator $M$ given by
\be
M = Q(h_0 -\mu_{\Phi} + 3 \langle N_\sigma\rangle g \Phi^2)Q
\ee
and an easy-to-reconstruct source $S$ term that we shall not write explicitly. Since $\Phi$ is the ground
energy solution of the Gross-Pitaevskii equation, the operator $M$ is positive and invertible 
(in the subspace orthogonal to $\Phi$). Whereas the explicit expressions of $M^{-1}$,
and thus of $M^{-1} S$, are unknown, in order to get the second term of (\ref{eq:derivtheta}) 
we fortunately only need 
\be
M^{-1} Q g \Phi^3 = - \partial_{\langle N_\sigma\rangle} \Phi,
\ee
as can be deduced from the derivative of (\ref{eq:gpe}) with respect to the particle number \cite{CastinDum}.
Then we get
\be
\int g \Phi^3 (\bar{\Xi}_\sigma+\bar{\Xi}_\sigma^*) =
\int (g Q \Phi^3) M^{-1} (-S) =
\int S \, \partial_{\langle N_\sigma\rangle} \Phi
\ee
All this leads to
\begin{multline}
\label{eq:thbar}
\!\!- t^{-1}\hbar \theta_\sigma(t) \underset{\mathrm{large}\, t}{\simeq} \mu_{\Phi}+
\left(N_\sigma\!-\!\langle N_\sigma\rangle-\!\!\int \overline{|\Lambda_\sigma^{(1)}|^2}\right) 
\partial_ {\langle N_\sigma\rangle} \mu_{\Phi} 
\\
+ \int \left[2\overline{|\Lambda_\sigma^{(1)}|^2}+\frac{1}{2}\left(\overline{\Lambda_\sigma^{(1)2}}
+\mbox{c.c.}\right)\right] \partial_{\langle N_\sigma\rangle} (\langle N_\sigma\rangle g\Phi^2) \\
- g \langle N_\sigma\rangle\!\!\!\int\!\!\!\!\int\!
\overline{(\Lambda_\sigma^{(1)}(\rr)\!+\!\mbox{c.c.})(\Lambda_\sigma^{(1)}(\rr')\!+\!\mbox{c.c.})} 
\Phi^3(\rr) \partial_{\langle N_\sigma\rangle} \Phi(\rr')
\end{multline}
where we used $\partial_{\langle N_\sigma\rangle} \mu_{\Phi} -\int g \Phi^4 = 2 \langle N_\sigma\rangle g 
\int \Phi^3 \partial_{\langle N_\sigma\rangle} \Phi$.
Remarkably (\ref{eq:thbar}) exactly reproduces the classical field version
of the full (not Gross-Pitaevskii) chemical potential 
$\mu$ of the gas at thermal equilibrium,
as given by Eq.~(103) of \cite{Mora}. This confirms the expectation that the condensate
phase evolves at the average pulsation $-\mu/\hbar$.

It remains to use (\ref{eq:asympt}) to obtain the squeezing parameter minimized over time.
As in our previous studies, the phase difference has the structure
\be
\theta_a(t)-\theta_b(t) \underset{\mathrm{large}\ t}{\simeq} -[N_a-N_b + D] 
\frac{\partial_{\langle N_a\rangle} \mu_{\Phi}}{\hbar} \, t
\label{eq:structure}
\ee
where $D$ collects the contributions that would be absent in the two-mode model realizing
the proposal of \cite{Ueda}. The time average in (\ref{eq:thbar}) suppresses the crossed terms that involve
two different Bogoliubov modes, since they oscillate in time. This leads to a single sum over modes:
\be
\label{eq:som_sim}
D = \sum_k d_k \left[\overline{|c_{ka}|^2}-\overline{|c_{kb}|^2}\right] 
= -\sum_k d_k [c_{ka}(0^-)B_k^*+ \mbox{c.c.}]
\ee
where we used (\ref{eq:init}) to transform $|c_{ka}(0^+)|^2-|c_{kb}(0^+)|^2$
and the fact that the quasi-particle occupation numbers $|c_{k\sigma}(t)|^2$
are constants of motion within Bogoliubov theory\footnote{This requires that the squeezing time
is shorter than the thermalisation time.
This holds in the weakly interacting limit \cite{prlsqueezing}.}. The general expressions for $d_k$
are easily deduced from (\ref{eq:thbar}). They look involved, but an inspired application
of the Hellmann-Feynman theorem to the Bogoliubov operator [of which $(u_k,v_k)$ is an eigenmode] gives
\be
\label{eq:dki}
d_k = \frac{\partial_{\langle N_\sigma\rangle} \epsilon_k}{\partial_{\langle N_\sigma\rangle} \mu_{\Phi}}
=\partial_{\mu_{\Phi}} \epsilon_k.
\ee
The simplicity of the result is understood from another expression of the gas
full chemical potential $\mu$ at thermal equilibrium, which is most rapidly
obtained within the microcanonical ensemble as in \cite{phase1}: $\mu$ is then the 
derivative of the energy $E$ with respect to the particle number $N_\sigma$ at fixed entropy;
since within Bogoliubov theory, $E = E_0+\sum_k \epsilon_k \langle |c_{k\sigma}|^2\rangle$,
and the occupation numbers $\langle |c_{k\sigma}|^2\rangle$ are constant at fixed entropy, 
we get (\ref{eq:dki}).

Finally, inserting (\ref{eq:structure}) in (\ref{eq:asympt}) and keeping the leading 
(first) order
in $\epsilon_{\rm Bog}$ gives for the classical field model\footnote{In particular, 
we used that $\langle (N_a-N_b)D\rangle
\approx N \epsilon_{\rm Bog}$, see Appendix D in \cite{italien},
so that its contribution, which appears squared in the numerator
of (\ref{eq:asympt}), is negligible as compared to $\langle D^2\rangle \approx N\epsilon_{\rm Bog}$.}:
\be
\label{eq:ana_class}
\xi^2_{\rm min} \stackrel{\mathrm{class}}{\simeq}  \frac{\langle D^2\rangle}{N} = \frac{1}{N} \sum_k d_k^2 \frac{k_B T}{\epsilon_k} 
(1+2 \langle v_k|v_k\rangle)
\ee
This expression is successfully compared to the numerical simulations (see symbols in fig.~\ref{fig:xi2class}a),
after a numerical diagonalisation of the Bogoliubov operator to obtain the eigenfunctions $(u_k,v_k)$ and
the eigenenergies $\epsilon_k$. 

\section{Extension to quantum fields}
The previous analytical developments have been performed within the classical field model, in order
to allow for a comparison with the numerical simulations. They can be quite directly
transposed to the quantum case, where the number conserving Bogoliubov theories were initially developed.
The condensate phases and particle numbers are now conjugate hermitian operators 
$[\hat{n}_\sigma, \hat{\theta}_{\sigma'}]=i\delta_{\sigma \sigma'}$, and the
classical fields become operators $\hat{\psi}_\sigma$ and $\hat{\Lambda}_\sigma$. 
Also $c_{k\sigma}, B_k$ and $D$ have quantum counterparts\footnote{In particular,
one has $[\hat{B}_k,\hat{B}_{k'}^\dagger]=\delta_{k k'}$ and
$\langle \hat{B}_k^\dagger \hat{B}_k\rangle=\langle v_k|v_k\rangle$.}.
The energy cut-off is no longer restricted to $k_B T$ and one can take the limit of a vanishing 
lattice spacing in the Bogoliubov results for $\xi_{\rm min}^2$, so that the notation $\int$ reduces to
a true spatial integral and one at last studies the case of 
trapped particles in real continuous space. We obtain
\be
\xi^2_{\rm min} \stackrel{\mathrm{quant}}{\simeq} \frac{\langle \hat{D}^2\rangle}{N}=
\frac{1}{N} \sum_k d_k^2 \left[
\frac{1+2 \langle v_k|v_k\rangle}
{e^{\epsilon_k/k_B T}-1} 
+\langle v_k|v_k\rangle\right]
\label{eq:res_quant}
\ee
The coefficients $d_k$ have the same expression in terms of the Bogoliubov modes
$(u_k,v_k)$ as in the classical field case, so that Eq.~(\ref{eq:dki}) still holds. 
As compared to (\ref{eq:ana_class}),
the occupation numbers of the Bogoliubov modes are now given by the Bose formula. There is
also a quantum term that subsists at $T=0$; the fact that $\xi_{\rm min}^2$
is non-zero for $T=0$ is due to the excitation of quasi-particles
by the $\pi/2$ pulse.

\section{Continuous spectrum limit}
Up to now, we have taken advantage of the large-$N$ limit (through the $\epsilon_{\rm size}$ expansion)
but we have not explicitly used the fact that the thermodynamic limit also corresponds to 
a vanishing single-particle level spacing, where the spectral sum in the quantum result
(\ref{eq:res_quant}) may be replaced
by an integral. This we now fully implement for an isotropic
\footnote{Modes of angular momentum $l$ are then degenerate;
their crossed terms in (\ref{eq:thbar}), not suppressed by
time average, are killed by spatial integration, so that the form
(\ref{eq:som_sim}) holds.}
harmonic trap $U(\rr)=m\omega^2 r^2/2$, where the semi-classical limit 
{\sl \`a la} WKB corresponds to 
$\hbar\omega/\mu_{\Phi}\to 0$ and can easily be taken thanks to integrability of the classical motion issued from 
the Bogoliubov equations \cite{Csordas}. 

For the condensate wavefunction, one takes the Thomas-Fermi approximation 
$W(r)\equiv 
\langle N_\sigma\rangle g \Phi^2(r)\simeq [\mu_{\Phi}-U(r)]\, Y[\mu_{\Phi}-U(r)]$ \cite{Leggett},  where $Y$ is the Heaviside
function; $\Phi$ then strictly vanishes at distances larger than the Thomas-Fermi radius $R$, and
one obtains 
\be
(\mu_{\Phi}/\hbar\omega)^3=
15 N (\pi/8)^{1/2}[\rho(\OO,0^-)a^3(0^-)]^{1/2}.
\ee
Due to rotational symmetry, the Bogoliubov modes are labeled by the radial $n\geq 0$, 
angular momentum $l\geq 0$ and azimuthal $m_l$ quantum numbers, so $k=(n,l,m_l)$.
The $m_l$-independent eigenenergies $\epsilon_{n,l}$ {(of $2l+1$ degeneracy)
are approximated by the Bohr-Sommerfeld rule} \cite{Csordas}
\be
\label{eq:BS}
\pi \hbar (n+1/2) = \int_{r_1}^{r_2} dr\, p_r(r,\epsilon_k,\mu_{\Phi})
\ee
where the positive classical radial momentum $p_r$, such that
\be
\frac{p_r^2(r,E,\mu)}{2m} = (E^2+W^2)^{1/2} -\left[\frac{\hbar^2 l^2}{2m r^2} + 2 W +U-\mu \right],
\ee
vanishes at the two classical turning points $r_1$ and $r_2$. Taking the derivative of
(\ref{eq:BS}) with respect to $\mu_{\Phi}$ gives
\be
d_k = \partial_{\mu_{\Phi}} \epsilon_k \simeq -\frac{\omega_{\rm cl}}{\pi} \int_{r_1}^{r_2} dr
\partial_{\mu} p_r (r,\epsilon_k,\mu_{\Phi}),
\ee 
where the angular frequency $\omega_{\rm cl}$ of the classical motion is given as usual
in classical mechanics by $\pi/\omega_{\rm cl}=\int_{r_1}^{r_2} dr \partial_E p_r(r,\epsilon_k,\mu_{\Phi})$.
Mode functions $u_k(\rr)$ and $v_k(\rr)$ are the product of spherical harmonics $Y_{l}^{m_l}(\theta,\phi)$,
and radial parts explicitly given by the WKB theory \cite{Csordas}, leading to
\be
1+2\langle v_k|v_k\rangle \simeq \frac{\omega_{\rm cl}}{\pi}  \int_{r_1}^{r_2} dr \frac{m}{p_r}.
\ee
{There exist two types of classical orbits in the trap, that are conveniently discussed in terms of  the reduced 
energy $\varepsilon\equiv E/\mu_{\Phi}$ and angular momentum
$j\equiv l \hbar\omega/(2\mu_{\Phi})$} \cite{Csordas}. A first type corresponds to
orbits that are purely out of the condensate, 
due to a large enough angular momentum (and centrifugal barrier),
and a not too high energy, $1<2j-1<\varepsilon < j^2$; in this case, the quasi-particles simply experience
the effective potential $U-\mu_{\Phi}$, so that $d_k\simeq -1$, $\omega_{\rm cl}=2\omega$ and $\langle v_k|v_k\rangle
\simeq 0$. The second type corresponds to {\sl mixed} orbits that cross the condensate boundary, 
that is $r_1 < R < r_2$, which corresponds to the parameter range $0<j^2<\varepsilon$.
Also in that case $d_k$ is negative (at variance with the homogeneous case), but it is larger than $-1$. 
This implies that $\xi_{\rm min}^2$ is always smaller than the non-condensed fraction as in \cite{prlsqueezing}.
Integration over the purely external orbits can be performed explicitly:
\be
\frac{\xi_{\rm min}^2}{[\rho(\OO,0^-)a^3(0^-)]^{1/2}} \simeq \frac{15\sqrt{\pi}}{2\sqrt{2}} 
\left(\frac{k_B T}{\mu_{\Phi}}\right)^3\!\! g_3(e^{-\mu_{\Phi}/k_B T}) + f_{\rm mix}
\label{eq:ana_quant}
\ee
where $g_\alpha(z)=\sum_{n\geq 1}z^n/n^\alpha$ is the Bose function. The mixed-orbit contribution
still involves a double integral, which is ultraviolet and infrared convergent:
\be
f_{\rm mix}=\frac{15\sqrt{2}}{\sqrt{\pi}} \int_0^{+\infty}d\varepsilon \int_0^{\varepsilon^{1/2}}
j dj \frac{I^2}{J} \Big[\frac{K}{J}\coth \frac{\varepsilon \mu_{\Phi}}{2 k_B T} -1\Big]
\ee
where the functions $I,J,K$ of $\varepsilon$ and $j$ are given in the appendix\footnote{
For $k_BT \gg \mu_{\Phi}$ external orbits dominate $f_{\rm mix}=O(k_BT/\mu_{\Phi})^2$.}.
This leads to the quantum result plotted as a solid line in fig.~\ref{fig:xi2class}a. 
{A local density approximation (LDA) for $\xi^2$ is plotted for comparison, although its use
to calculate a spin variance is {\it a priori} not justified. We use
$(N\xi^2)_{\rm LDA}=\int d^3r \, (\rho \xi^2)_{\rm hom}[\mu_{\rm hom}=\mu_{\Phi}-U({\rr})]$. The quantity
$(\rho \xi^2)_{\rm hom}$ for the homogeneous system is deduced from Eq.(\ref{eq:res_quant})
if $\mu_{\rm hom}>0$ and it is given by the ideal gas density elsewhere ($\xi_{\rm hom}^2=1$). For $k_BT\gg\mu_{\Phi}$,
$\xi^2$ is asymptotic to the non-condensed fraction, hence the success of the LDA.}

\section{Conclusion}
We show that spin squeezing driven by interactions in trapped bimodal atomic condensates 
takes a finite optimal value
in the thermodynamic limit, a value that we determine analytically for weak interactions. Our theory 
applies for classical (\ref{eq:ana_class}) and quantum fields (\ref{eq:res_quant}),
and for arbitrary trap geometries. 
It indicates
that a large metrologic gain is possible at finite temperature in realistic experimental conditions:  
for $k_BT\simeq\mu_{\Phi} \simeq 10 \hbar \omega$, and assuming that one million is a large enough atom number 
to reach the thermodynamic limit, we predict a signal-to-noise increase by a factor $1/\xi\simeq 30$.

\acknowledgements 
A.S.\ and E.W.\ thank the CNRS/PAN collaboration. A.S. thanks ACUTE and QIBEC projects.
E.W.\ acknowledges NSC grant 2011/03/D/ST2/01938.

\section{Appendix}

With the change of variable $1-r^2/R^2=\varepsilon \sinh z$, and then $X=e^{\pm z}$, the
semi-classical integrals for mixed orbits can be evaluated. Setting $J=\pi \omega/\omega_{\rm cl}=J_1+J_2$,
where $J_1$ ($J_2$) is the in(out)-condensate contribution, one finds:
\bea
J_1 &=& \frac{\varepsilon/\sqrt{2}}{(2j^2+\varepsilon^2)^{1/2}} \arccos 
\left[\frac{2j^2+\varepsilon^2-\varepsilon}{\varepsilon (2j^2+\varepsilon^2+1)^{1/2}}\right] \\
J_2 &=& \frac{1}{2} \arccos \frac{1-\varepsilon}{[(1+\varepsilon)^2-4 j^2]^{1/2}}
\eea
Setting $I=-J \partial_{\mu_{\Phi}} \epsilon_k=I_1+I_2$, one finds $I_2=J_2$ and
\be
I_1=-\frac{1}{\sqrt{2}} \, \mathrm{arccosh}\, \frac{1+\varepsilon}{(2j^2+\varepsilon^2+1)^{1/2}}
\ee
Setting $K=J (1+2\langle v_k|v_k\rangle)=K_1+K_2$, one finds $K_2=J_2$ and
\be
K_1=\frac{\varepsilon(J_1+\sqrt{\varepsilon-j^2})}{2(\varepsilon^2+2j^2)}-\frac{I_1}{2} 
\ee


\begin{thebibliography}{99}

\bibitem{Santarelli}
G. Santarelli {\sl et al}., Phys. Rev. Lett. {\bf 82}, 4619 (1999).

\bibitem{Ueda}
M. Kitagawa, M. Ueda, Phys. Rev. A {\bf 47}, 5138 (1993).

\bibitem{Wineland}
D.J. Wineland {\sl et al.}, Phys. Rev. A {\bf 50}, 67 (1994).

\bibitem{Polzik}
A. Louchet-Chauvet  {\sl et al}., 
NJP {\bf 12}, 065032 (2010).

\bibitem{Vuletic}
I.D. Leroux, 
Phys. Rev. Lett. {\bf 104}, 073602 (2010).

\bibitem{NatureSorensen}
A. S\o rensen {\sl et al.}, Nature {\bf 409}, 63 (2001).

\bibitem{manips}
C. Gross {\sl et al.}, Nature (London) {\bf 464}, 1165 (2010);
M.F. Riedel {\sl et al.}, {\sl ibid.}, 1170 (2010).

\bibitem{LiYun}
Yun Li {\sl et al}., 
Phys. Rev. Lett. {\bf 100}, 210401 (2008).

\bibitem{prlsqueezing}
A. Sinatra {\sl et al.}, 
Phys. Rev. Lett. {\bf 107}, 060404 (2011).

\bibitem{italien}
A. \!Sinatra \!{\sl et al}., \!
Eur. \!Phys. \!J. \!Special \!Topics {\bf 203}, \!87 \!(2012).

\bibitem{frontiers}
A. Sinatra {\sl et al}., 
Front. Phys. {\bf 7}, 86 (2012); G. Ferrini {\sl et al}.,
Phys. Rev. A {\bf 84}, 043628 (2011).

\bibitem{Isella}
L. Isella, J. Ruostekoski, Phys. Rev. A {\bf 74}, 063625 (2006).

\bibitem{Opanchuk}
B. Opanchuk {\sl et al}., 
EPL {\bf 97}, 50003 (2012).

\bibitem{Lukin}
E.G. \!Dalla \!Torre {\sl et al}., \!
Phys. \!Rev. \!Lett. \!{\bf 110}, \!120402 \!(2013).

\bibitem{spsq_dyn} Yun Li {\sl et al}., Eur. Phys. J.B {\bf 68}, 365 (2009).

\bibitem{Cartago}
A. Sinatra {\sl et al}., 
J. Phys. B {\bf 35}, 3599 (2002).

\bibitem{Gardiner}
C.W. Gardiner, Phys. Rev. A {\bf 56}, 1414 (1997).

\bibitem{CastinDum} 
Y. Castin, R. Dum, Phys. Rev. A {\bf 57}, 3008 (1998).

\bibitem{phase1}
A. Sinatra {\sl et al}., 
Phys. Rev. A {\bf 75}, 033616 (2007).

\bibitem{Sorensen}
A. S\o rensen, Phys. Rev. A {\bf 65}, 043610 (2002).


\bibitem{Mora}
C. Mora, Y. Castin, Phys. Rev. A {\bf 67}, 053615 (2003).

\bibitem{Csordas}
A. Csord\'as {\sl et al}., 
Phys. Rev. A {\bf 56}, 5179 (1997).

\bibitem{Leggett}
V.V. Goldman {\sl et al}., 
Phys. Rev. B {\bf 24}, 2870 (1981).

\end{thebibliography}
\end{document}